\documentclass[%
 reprint, amsmath,amssymb, aps,showpacs]{revtex4-1}

\usepackage{graphicx}
\usepackage{epsfig}
\begin{document}


\title{ Thermoelectric transport in the topological phase due to 
the coexistence of superconductivity and spin-density-wave}

\author{Amit Gupta}

\author{Debanand Sa}%
\affiliation{%
Department of Physics, Banaras Hindu University, Varanasi-221 005
  \\}%
  
\date{\today}

\begin{abstract}
We study the thermoelectric transport in two dimensional topological system which 
has coexistence of superconductivity(SC) and spin-density wave(SDW). The SC is presumed
to be of $d_{x^2-y^2}+(p_x + i p_y) $  type whereas the SDW order parameter is of $BCS$ symmetry. The Hamiltonian describing such a coexistence phase is shown to have topological phase in addition to the conventional one. The transport properties in such topological system have two distinct contributions: (i) the surface/edge and (ii) the bulk. 
The competition between the surface/edge versus the bulk transport is analyzed in 
different parameter regimes and the possibility of enhancing the figure of merit is discussed.
\begin{description}
\item[PACS numbers]{74.25.fg, 75.30.Fv, 73.43.-f, 74.20.Mn}
\end{description}
\end{abstract}

\maketitle



In last few years, a new field in condensed matter systems has emerged which is called 
"Topological insulators and superconductors"\cite{hasan10,qi11}. It is based on the realization that the spin-orbit interaction in materials can lead to such electronic 
phases\cite{kane05,sheng05,bernevig06,hughes06} which has been observed in real materials\cite{konig08,moore10,qi10}. 
A topological insulator, similar to an ordinary insulator has a bulk energy gap 
separating the highest occupied electronic band from the lowest empty band\cite{moore91,hatsugai93,nayak96,fradkin98}.  It is closely related to the two-dimensional integer quantum Hall states\cite{thouless82} 
which have unique edge states. The surface of such insulators however, necessarily has 
gapless states (edge states) that are protected by time reversal symmetry. These states 
are conducting with properties unlike any other known one-dimensional or two-dimensional 
electronic systems. These states are predicted to have special properties which are 
thought to be useful for applications ranging from spintronics to quantum 
computation\cite{nayak08}. 
Due to such technological importance the field of topological systems has grown rapidly 
over the years. Since one of the compelling criteria of a material to be topological 
is that the system ought to have an energy gap, the topological concept extends over to 
any gapped systems such as superconductors and superfluids 
etc.\cite{read00,ivanov01,stern04}. Recently, a possibility of a topological phase in 
the coexistence phase of superconductivity and spin-density wave has been discussed\cite{lu14,tanmoy13,gupta15}. The nature of such topological phase depends 
on the symmetry as well as the amplitudes of both the order parameters. 

Recently, the photoemission and the scanning tunneling spectroscopy have become an 
important experimental tool in identifying the surface states in topological systems 
but the signature of such states in transport measurements is still under debate. The focus 
of experimental studies thus has shifted to the separation of bulk conduction from the 
surface conduction. Recently, there is an attempt to formulate the thermoelectric 
transport in topologial insulators theoretically\cite{takahashi10}. Thermoelectric 
transport is basically conversion of heat to energy. The efficiency of such energy 
conversion depends on the "figure of merit" of the material which is defined as, 
$ZT=\frac{\sigma S^2 T}{\kappa}$\cite{goldsmid64}, 
where $T$ is the temperature, $S$ is the Seebeck coefficient and $\sigma$ and $\kappa$ 
are respectively the electrical and thermal conductivities. In order to achieve a high 
$ZT$, the system should be a good electronic conductor whereas a bad lattice conductor. 
Further, the material low-dimensionality also have large $S$ and have high efficiency 
due to their peaked density of states\cite{hicks93}. Despite such proposals it was very 
hard to discover good thermoelectric materials in the past. This is due to the fact that 
$\sigma$, $S$ and $\kappa$ cannoot be independently controlled, namely, a material with 
large electrical conductivity $\sigma$ has large thermal conductivity $\kappa$. 
The discovery of topological systems
gives some hope on the above issues due to the fact that the edge state conduction 
remains good whereas the phonon conduction is suppressed. The edge states are 
one-dimensional which satisfies the low dimensionality\cite{dress93} criteria mentioned above. The topological phenomena which has already been observed in materials such as 
$Bi_{1-x}Sb_x$\cite{hsieh08}, $Bi_2Se_3$\cite{xia09} and $Bi_2Te_3$\cite{chen09} are 
known as good thermoelectric materials.  

In this communication, we study the thermoelectric transport in two-dimensional 
topological system which has coexistence of superconductivity(SC) and spin-density 
wave(SDW). The symmetry of SC is considered to be of $d_{x^2-y^2}+(p_x + i p_y)$  
type whereas the SDW order parameter is of $BCS$ symmetry. The Hamiltonian having 
such a structure is shown to have topological coexistence phases in addition to the conventional one. The thermoelectric transport in such topological system is discussed 
with respect to two distinct contributions: (i) the surface and (ii) the bulk. The competition among the surface versus the bulk transport is analyzed in different 
parameter regimes and the possibility of enhancement of figure of merit is discussed 
in these systems.


In the present work, we consider the coexistence of SDW and d-wave superconductivity 
which can generate a triplet and non-zero center of mass superconducting order parameter. 
We, thus start with a Hamiltonian on a $ 2 $D  square lattice\cite{gupta15,dsa15} as,  

\begin{widetext}
\begin{eqnarray}
{\cal H}=
\sum_{k, \sigma} \xi_{k}c^{\dagger}_{k,\sigma}c_{k,\sigma}+\frac{U}{N•}\sum_{k,k'}c^{\dagger}_{k,\uparrow}c_{k+Q,\uparrow}c^{\dagger}_{k'-Q,\downarrow}
c_{k',\downarrow} 
+ \sum_{k,k'}V^{1}(k,k')c^{\dagger}_{k,\uparrow}c^{\dagger}_{-k,\downarrow}c_{-k',\downarrow}c_{k',\uparrow}
\nonumber\\
 +\sum_{k,k'}V^{2}(k,k')c^{\dagger}_{k,\uparrow}c^{\dagger}_{-k-Q,\downarrow}
 c_{-k'-Q,\downarrow}c_{k',\uparrow}. 
\end{eqnarray}
\end{widetext}

\noindent Here, $ \xi_{k} $ is the bare dispersion due to the tight binding 
approximation on a 2D square lattice, $U$ is the on-site Coulomb interaction, 
$ V^{1,2} $ are the pairing strengths for $d$-wave and $p$-wave superconductivity 
and $N$ is the number of sites. Also,  
$ \xi_{k} =-2t(\cos k_{x} +\cos k_{y}) - 4t'\cos k_{x}\cos k_{y} -\mu $ and  
$c^{\dagger}_{{{k}}\sigma}$ ($c_{{{k}}\sigma}$) denotes creation (annihilation) operator 
of the  electron with spin $ \sigma=(\uparrow,\downarrow)$  at ${\textbf{k} }=(k_x,k_y)$. Here, $\sum_{k}^{'}$ is the sum of $k$ over the reduced Brillouin zone (RBZ). 
We express the wave-vector $k$ in units of  $ \frac{\pi}{a} $, with '$a$' the lattice parameter of the underlying square lattice. ${\bf Q} = (\pi, \pi)$ is the SDW nesting 
vector in 2D. We assume here a commensurate 
SDW so that $ {\bf {k + Q}} ={\bf {k-Q}} $. The staggered spin magnetization is defined as $M_0 = -\frac{U}{N}\sum_{k, \sigma}\sigma<c^{\dagger}_{k+Q,\sigma}c_{k,\sigma}>$. Since  
the discussion would be about three order parameters below, the crystal symmetry are
such that the commutator of any two of them should give the third one. Thus, 
if $ V^{1}$ is assumed to be of singlet d-wave symmetry, the SDW state guarantees that 
$ V^{2}$  should be of triplet type. So we get the singlet 
interaction $V^{1}_{k,k'} =V_{0}^{1} s_{k} s_{k'}$ and $V^{2}_{k,k'} =V_{0}^{2} p_{k} p_{k'}$, where $s_{k} =\frac{1}{2}(\cos k_{x} -\cos k_{y}) $ and 
$p_{k} =\sin k_{x} + i\sin k_{y}$. 
We further assume that $V^{1,2}$ are attractive. The SC order parameters are defined as, 
for singlet state, $\bigtriangleup^{1}_{k'} = \bigtriangleup^{1}_{0} s_{k'} = V_{0}^{1} s_{k'}\sum_{k}s_{k}<c^{\dagger}_{k,\uparrow}c^{\dagger}_{-k,\downarrow}> =V_{0}^{1} s_{k'}  \sum_{k}^{'}s_{k}<c^{\dagger}_{k+Q,\uparrow}c^{\dagger}_{-k+Q,\downarrow}>$.  On the other hand, the triplet order is $\bigtriangleup^{2}_{k'} = \bigtriangleup^{2}_{0} p_{k'} = V_{0}^{2} p_{k'}\sum_{k}p_{k}<c^{\dagger}_{k+Q,\uparrow}c^{\dagger}_{-k,\downarrow}>= \bigtriangleup_{0}^{2}(\sin k_{x} + i\sin k_{y})=\bigtriangleup_{1,k}^{2}+ i\bigtriangleup_{2,k}^{2} $ while $\bigtriangleup^{2\ast}_{k'} = V_{0}^{2} p_{k'}\sum_{k}p_{k}<c^{\dagger}_{k,\uparrow}c^{\dagger}_{-k+Q,\downarrow}>$. 
%
%
Defining $ \xi_{k}^{+}=- 4t'\cos k_{x}\cos k_{y} -\mu $ and
$ \xi_{k}^{-}=-2t(\cos k_{x} +\cos k_{y}) $ and employing the nesting property
in the band dispersion i.e. 
$\xi_{k+Q}^{+}= \xi_{k}^{+} $, $\xi_{k+Q}^{-}= -\xi_{k}^{-}$ and also the order parameters  
$\bigtriangleup^{1}_{k+Q} 
=-\Delta_{0}^{1}(\frac{\cos k_{x}a-\cos k_ya}{2}) 
=-\bigtriangleup^{1}_{k}$ and 
$ \bigtriangleup^{2}_{k+Q}=-\bigtriangleup^{2}_{k} $,  
the Hamiltonian in the momentum space can be expressed as, 
${\cal H}=\sum_{k}\psi^{\dagger} _{k} {\cal H}({k})\psi_{k}$ 
where the four-component spinor $\psi_{k}$ is,  
$\psi^{\dagger}_{k}=(c_{k\uparrow}^{\dagger},c_{-k-Q\downarrow},
c_{-k\downarrow}^{\dagger},c_{k+Q\uparrow})$.   
Thus, the Hamiltonian matrix  $  {\cal H}({k}) $ in this basis is written as, \\
 
\begin{equation}
\label{eq.2}
{\cal H}({k})= \left( \begin{array}{cccc}
\xi_{k}^{+}+ \xi_{k}^{-}& \bigtriangleup^{2}_{k}  & \bigtriangleup^{1}_{k} &  M_{0}\\
\bigtriangleup^{2\ast}_{k} & -\xi_{k}^{+}+ \xi_{k}^{-} & M_{0}  & -\bigtriangleup^{1}_{k}\\
\bigtriangleup^{1}_{k} &  M_{0} & -(\xi_{k}^{+}+ \xi_{k}^{-})  & -\bigtriangleup^{2\ast}_{k} \\
M_{0} & -\bigtriangleup^{1}_{k}& -\bigtriangleup^{2}_{k} & \xi_{k}^{+}- \xi_{k}^{-}
\end{array} \right).
\end{equation} 

In what follows, we study the energy spectrum of the above Hamiltonian. 
The Hamiltonian (eqn.(3)) is diagonalized and the quasiparticle spectrum is obtained as,

\begin{widetext}
$E_{\pm,\pm}(k) 
=\pm\sqrt{\xi_{k}^{+2} +\xi_{k}^{-2}+(\bigtriangleup^{1}_{k})^2 
+ \mid \bigtriangleup^{2}_{k}\mid^{2}+M_{0}^{2}\pm 2 \sqrt{\xi_{k}^{-2}\mid \bigtriangleup^{2}_{k}\mid^{2} +
(\bigtriangleup^{2}_{1,k} \bigtriangleup^{1}_{k}-M_{0}\xi_{k}^{+})^{2}+\xi_{k}^{+2} 
\xi_{k}^{-2} }}$.
\end{widetext} 

\begin{figure}
\includegraphics[scale=0.50]{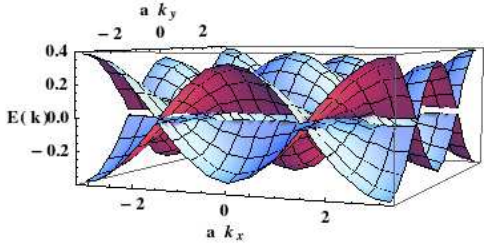} 
\caption{Energy spectra  $ E_{\pm,+}(k)$, corresponding to 
coexistence of SC order parameters $ d_{x^{2}-y^{2}}+(p_{x}+ ip_{y})$ and that of 
the SDW order  parameter showing fully gapped spectrum. For 
illustration, here, we have chosen $t'=-0.3 z $, $M_0=0.075 z$, 
$ \bigtriangleup_{0}^{1}=\bigtriangleup_{0}^{2}=0.005 z$, ($z=0.30$ eV). }
\label{fig.1}
\end{figure}

\noindent It is obvious that the energy spectrum is fully gapped as shown in Fig. 1 
and the gap closes only 
when the rhs of the above equation vanishes. A straight forward calculation provides 
a condition where the gap closes at points $k=(0,0)$ and $(\pi, \pi)$ is 

\begin{equation} 
16 t^2 + M_0^2=(4t'+\mu)^2. 
\end{equation} 

\noindent Thus, one can find a topologically trivial and non-trivial regions with respect 
to these parameters. Since the critical phase lines(for $t=0$) are determined by the condition 
$M_0^2=(4t'+\mu)^2$ the phase becomes topological in the region where $M_0^2<(4t'+\mu)^2$ whereas it is trivial for $M_0^2>(4t'+\mu)^2$. The details of these work have been 
discussed in a recent paper\cite{dsa15}.   
 
In order to study thermoelectric transport in these system, we consider a 2-D sample as 
a ribbon geometry where the ribbon width is taken to be very narrow. This is due to the 
fact that the edge states can have comparable contribution as compared with the bulk. 
Using a linear response theory\cite{mahan00}, the electric current $j$ and thermal 
current $w$ which are coupled, are written as 

\begin{equation}
\left(\begin{array}{c}  
j/q \\ w \end{array}\right)=
\left(\begin{array}{cc} L_{0} & L_{1} \\ L_{1} & L_{2} \end{array}\right)
\left(\begin{array}{c}
-\frac{\mathrm{d}\mu}{\mathrm{d}x} \\ -\frac{1}{T}\frac{\mathrm{d}T}{\mathrm{d}x} \end{array}\right),
\end{equation}

\noindent where $q$ is the electron charge $-e$, $\mu$ is the chemical potential. Here, 
the field used are the electric field and the thermal gradient. The electric and thermal transport coefficients are obtained as, 

\begin{eqnarray}
&&\sigma=e^2L_0,\ \  \ S = -\frac{1}{eT}\frac{L_1}{L_0},\ \ \ 
\kappa_e = \frac{1}{T}\frac{L_0 L_2 - L_1^2}{L_0},\nonumber\\
&&ZT = \frac{L_1^2}{L_0 L_2 - L_1^2 + \kappa_L T L_0},
\end{eqnarray}

\noindent where $\kappa_e$ is the electron thermal conductivitiy and $\kappa_L$ is phonon thermal conductivity. Here, $L_\nu$'s are the correlation functions determing the thermoelectric transport coefficients.

Since the topological systems have distinct edge and bulk states, we consider the 
transport due to both independently. We first consider the edge transport only. 
To describe the coherent transport of the edge states, we use the Landauer formula.
The edge states are assumed to be perfectly conducting and the transmission coefficient $T(E)$ is taken as unity. This is true when the electron energy is within the bulk gap 
($-\Delta < E < \Delta $), $\Delta$ being the effective gap. 
The energy here is measured from the bottom 
of the conduction band. While carrying out the calculation of the transport coefficients, 
we consider the bottom of the bulk conduction band and neglect the valence band such that 
we deal only with the edge states. We further, restrict the chemical potential $\mu$ to be within the gap. The edge state correlation fuction $L_\nu$ is given by

\begin{align}
L^{{\rm e}}_{\nu}(\mu) =\frac{2\ell}{sh}\int \mathrm{d}E T(E)(E-\mu)^{\nu} \left(-\frac{\partial f}{\partial E} \right),
\end{align}

\noindent where the suffix '$e$' means the edge transport, $h$ is the Planck constant and  
$\ell$ and $s$ respectively are the length and the cross-section of the sample, the 
factor 2 comes from the two gapless channels of the 2D system.
The integral in the above correlation function can be rewritten in a dimensionless form as 

\begin{eqnarray}
L_{\nu}^{e}(\bar\mu) 
=
\frac{2\ell}{s h}{(k_B T)^{\nu}}
\int_{-\bar\Delta}^{\bar\Delta} dy      
(\bar{y}-\bar\mu)^{\nu}
\frac{e^{(\bar{y}-\bar\mu)}}{({e^{(\bar{y}-\bar\mu)}+1})^2},
\label{eq:xx} 
\end{eqnarray}  

\noindent where $\bar{y}=\frac{E}{k_B T}$, $\bar\Delta=\frac{\Delta}{k_B T}$, $\bar\mu=\frac{\mu}{k_B T}$ and $\beta=\frac{1}{k_B T}$, $T$ being the temperature. 
Using this edge-state correlation functions the thermoelectric coefficients 
are calculated. It is noticed that the edge contribution to the thermoelectric 
coefficients such as the figure of merit (ZT) is independent of the system size $\ell$ 
and $s$ eventhough the edge correlation functions are directly proportional to $\ell$ 
and inversely proprtional to the cross-sectional area $s$. Further, ZT is 
unusually large and exceeds unity when the chemical potential is in the bulk band. This 
is due to the fact that when $\mu$ is in the bulk band, the correlation functions $L_0$ 
has only exponential dependence on $\mu$ but $L_1$ and $L_2$ have combined exponential 
and algebraic dependence. On the contrary, when the chemical potential is in the bulk gap, 
ZT becomes exponentially small ($\mu\rightarrow 0$) which is in agreement with the figures 
(see Fig. 2 and Fig. 3).  
 
Next, we consider the bulk thermoelectric transport. The bulk transport are calculated 
using the Boltzmann transport theory. We assume that the relaxation time $ \tau_e $ is constant. Since these transports are valid within the inelastic scattering length, we 
regard the inelastic scattering length as the effective system size. We use only the 
upper two subbands because there is a large gap between the upper and lower subbands due 
to the narrow-ribbon confinement. The bulk correlation function $L_{\nu}$ is written as  

\begin{align}
L^{{\rm b}}_{\nu}(\mu) =\int \mathrm{d}E (E-\mu)^{\nu} \left(-\frac{\partial f}{\partial E} \right) D(E) \tau_e v^2,
\end{align}

\noindent where the suffix '$b$' means the bulk transport and $\tau_e$ is the relaxation 
time which is assumed to be constant. Also $D(E)$ and $v$ respectively are the bulk DOS 
and the velocity. The above integral can be changed to integral over the momentum 
variables as
 
\begin{eqnarray}
L_{\nu}^{b}(\mu) &=&\frac{\tau_e}{c}\int\frac{d^{2}k}{(2\pi)^{2}}
\left(\frac{\partial E_{++}(k)}{\hbar\partial k_{x}}\right)^{2}(E_{++}(k)-\mu)^{\nu}\nonumber\\
&& (-\frac{\partial f(E_{++}(k)-\mu)}{\partial E_{++}})+(E_{++}\rightarrow E_{+-}). 
\label{eq:xx} 
\end{eqnarray} 

\noindent Using dimensionless variables namely $k_x a=y_1$ and $k_y a=y_2$, the above equation is written as 

\begin{eqnarray}
L_{\nu}^{b}(\bar\mu) 
=\frac{\tau_e}{c\hbar^2}{(k_B T)^{\nu+1}}
\int_{-\pi}^{\pi} dy_1\int_{-\pi}^{\pi} dy_2      
(\frac{\partial \bar{E}_{+}(y_1,y_2)}{\partial y_1})^2
\nonumber \\
(\bar{E}_{+}(y_1,y_2)-\bar\mu)^{\nu}
\frac{e^{(\bar{E}_{+}(y_1,y_2)-\bar\mu)}}{({e^{(\bar{E}_{+}(y_1,y_2)-\bar\mu)}+1})^2}. 
\label{eq:xx} 
\end{eqnarray}  

\noindent The variable $\bar{E}_{+}$ in the above equations is written as, $\bar{E}_{+}=\frac{E_{+}}{k_B T}$ which is functions of $y_1=k_x a$ and $y_2=k_y a$.  
Using the parameters in the energy dispersion, the above correlation functions are 
calculated. It is noted that all the bulk correlations $L_{\nu}$ behave algebraically 
when the chemical potential is in the bulk band whereas they have combinations of 
algebraic and exponential dependence when $\mu$ is in the bulk gap. Meanwhile, 
the figure of merit is larger when the chemical potential is away from the band edge.

Since the actual thermoelectric transport coefficients involve both the edge and bulk correlation functions, we use $L_{\nu}=L_{\nu}^e +L_{\nu}^b$ and compute the transport properties. It has already been mentioned in the earlier section that the thermoelectric transports in the topological systems are due to the competition between the edge and 
the bulk contributions. As has already been discussed above, the relative magnitudes 
of both the edge and the bulk transports for different chemical potentials are different, 
the figure of merit ZT is also different. When the chemical potential is in the bulk band, 
ZT from the edge becomes larger eventhogh the number of edge carriers are 
exponentially small. This is due to the fact that that the edge transport is overridden 
by the bulk contributions and a high ZT from the edge never appears. In a similar way, ZT from the bulk becomes larger when the chemical potential is in the bulk gap. In this case, there are very few bulk carriers whereas there are some edge carriers for which the edge carriers dominate and hence ZT is suppressed. This is a clear indication that in topological system the edge and the bulk transport compete with each other suppressing the total ZT.\\

In order to compute the transport properties, we consider the following parameters: the hopping integral is taken to be zero and the next nearest neighbour hopping integral as $t'=-0.3 z$. The SC and SDW order parameters respectively are considered to be $\Delta_0^{1}=\Delta^{2}_{0}=0.005 z$, $M_0=0.075 z$ where the parameter $z$ taken to be 
$z=0.30 eV$. The computation is performed at a temperature $T=20K$. The relaxation time 
is taken to be $\tau_e=10^{-13} sec$. The phonon thermal conductivity is taken to be 
constant which is $\kappa_L=0.5 W {m}^{-1}{K}^{-1}$. The effective system size $\ell$ is assumed to be $1\mu m$ and the cross-sectional area $s$ is taken as $10 nm\times 0.1 nm$.  All these parameters are taken in an adhoc basis and the thermoelectric coefficients are computed. The results are shown in Fig. 2. and Fig. 3. It is obvious from Fig. 2 and Fig. 3 
that the edge contribution to $\sigma$, $\kappa_e$ and $ZT$ is much higher as compared 
to the bulk. On the contrary, the Seebeck coefficient has comparable contribution from both, 
a positive contribution from edge whereas a negative contribution from bulk. Thus, there 
is a partial cancellation due to opposite charge carriers in the edge and the bulk. This 
results a peak structure in the Seebeck coefficient near the band edge which makes the 
figure of merit large. 

Due to the edge-bulk competition, there is a possibility that ZT could 
be maximum when the chemical potential is near the band edge. In such a case, the bulk 
conduction is dominant at high temperature. On lowering temperature, the bulk-edge 
cross-over takes place and there is a possibility that ZT might be higher as shown in 
Fig.2 and Fig.3. The edge 
current in these system provides ballistic transport but such transport crucially 
depends on the inelastic scattering length $\ell$ of the edge states because they loose 
their coherence due to inelastic scattering. At high temperature, $\ell$ is short and 
the bulk transport is dominant but at low temperature, $\ell$ is long and the edge 
transport is dominant. This might be the reason why an enhanced figure of merit results 
at low temperature. 

In conclusion, we summarize the main findings of the present work. 
We study the thermoelectric transport in a two-dimensional 
topological system which has coexistence of superconductivity(SC) and spin-density 
wave(SDW). The SC is presumed to be of $d_{x^2-y^2}+(p_x + i p_y)$  type 
whereas the SDW order parameter is of $BCS$ symmetry. The Hamiltonian having 
such a structure is shown to have topological coexistence phases in addition to the conventional one. The thermoelectric transport in such topological system is discussed 
with respect to two distinct contributions: (i) the surface and (ii) the bulk. The competetion among the surface versus the bulk transport is analyzed in different 
parameter regimes and the possibility of enhancement of figure of merit is discussed 
in these systems.\\


\section{Acknowledgements}    
Financial supports from CSIR, India are gratefully 
acknowledged.


\begin{thebibliography}{9}

\bibitem{hasan10} 
M. Z. Hasan and C. L. Kane, Rev. Mod. Phys. {\bf 82}, 3045 (2010).

\bibitem{qi11} 
Xiao-Liang Qi and Shou -Cheng Zhang, Rev. Mod. Phys. {\bf 83}, 1057 (2011).

\bibitem{kane05} 
C. L. Kane and E. J. Mele, Phys. Rev. Lett. {\bf 95}, 146802 (2005). 

\bibitem{sheng05} 
L. Sheng, D. N. Sheng, C. S. Ting and F. D. M. Haldane, Phys. Rev. Lett. {\bf 95}, 
136602 (2005). 
 
\bibitem{bernevig06}  
B. A. Bernevig and S. -C. Zhang, Phys. Rev. Lett. {\bf 96}, 106802 (2006). 

\bibitem{hughes06} 
B. A. Bernevig, T. L. Hughes and S. -C. Zhang, Science {\bf314}, 1757 (2006). 

\bibitem{konig08}
M. K\"{o}nig, S. Wiedmann, C. Br\"{u}ne, A. Roth, H. Buhmann, L. W. Molenkamp, X. -L. Qi,  
and S. -C. Zhang , Science {\bf 318}, 766 (2008).

\bibitem{moore10}
J. E. Moore, Nature (London) {\bf464}, 194 (2010). 

\bibitem{qi10}
X. L. Qi and S. C. Zhang, Phys. Today {\bf 63}, 33 (2010).

\bibitem{moore91} 
G. Moore and N. Read, Nucl. Phys. B {\bf 360}, 362 (1991). 

\bibitem{hatsugai93} 
Y. Hatsugai, Phys. Rev. Lett. {\bf 71}, 3697 (1993). 

\bibitem{nayak96} 
C. Nayak and F. Wilczek, Nucl. Phys. B {\bf 479}, 529 (1996).
 
\bibitem{fradkin98} 
E. Fradkin, C. Nayak, A. Tsvelik and F. Wilczek, Nucl. Phys. B {\bf 516}, 704 (1998). 

\bibitem{thouless82}
D. J. Thouless, M. Kohmoto, M. P. Nightingale and M. den Nijs, Phys. Rev. 
Lett. {\bf 49} 405 (1982).

\bibitem{nayak08} C. Nayak, S. H. Simon, A. Stern, M. Freedman and S. Das Sarma, 
Rev. Mod. Phys. {\bf 80}, 1083 (2008).

\bibitem{read00} 
N. Read and  D. Green, Phys. Rev. B {\bf 61}, 10267 (2000). 

\bibitem{ivanov01} 
D.A. Ivanov, Phys. Rev. Lett. {\bf 86}, 268 (2001). 

\bibitem{stern04} 
A. Stern, F. von Oppen and E. Mariani, Phys. Rev. B {\bf 70}, 205338 (2004).

\bibitem{lu14} Yuan-Ming Lu, Tao Xiang and Dung-Hai Lee, Nature Phys. {\bf 10}, 634 (2014).

\bibitem{tanmoy13} Tanmoy Das, arXiv:1312.0544 v1 [cond-mat.supr-con] 02 Dec. (2013).

\bibitem{gupta15} Amit Gupta and Debanand Sa, Solid State Commun. {\bf 203}, 41 (2015).

\bibitem{dsa15} Amit Gupta and Debanand Sa, arXiv:1504.04969 v1 [cond-mat.supr-con] 
17 March (2015).

\bibitem{takahashi10} R. Takahashi and S. Murakami, Phys. Rev. B {\bf 81}, 161302 (2010). 

\bibitem{goldsmid64}
H. J. Goldsmid, {\it Thermoelectric Refrigeration} (New York, Plenum) (1964).

\bibitem{hicks93} L. D. Hicks and M. S. Dresselhaus, Phys. Rev. B {\bf 47}, 12727 (1993). 

\bibitem{dress93} L. D. Hicks and M. S. Dresselhaus, Phys. Rev. B {\bf 47}, 16631 (1993). 

\bibitem{hsieh08} D. Hsieh, D. Qian, L. Wray, Y. Xia, Y. S. Hor, R. J. Cava and M. S. Hasan, Nature {\bf 452}, 970 (2008).

\bibitem{xia09} Y. Xia, D. Qian, D. Hsieh, L. Wray, A. Pal, H. Lin, A. Bansil, D. Grauer, 
Y. S. Hor, R. J. Cava and M. Z. Hasan, Nature Phys. {\bf 5}, 398 (2009).

\bibitem{chen09} Y. L. Chen, J. G. Analytis, J. -H. Chu, Z. K. Liu, S. -K. Mo, X. L. Qi, 
H. J. Zhang, D. H. Lu, X. Dai, Z. Fang, S. C. Zhang, I. R. Fisher, Z. Hussain and 
Z. -H. Shen, Science {\bf 325}, 178 (2009).

\bibitem{mahan00}
G. D. Mahan, {\it Many-Particle Physics} (New York, Plenum) (2000).


\end{thebibliography}

\begin{figure}[h]
\begin{center}$
\begin{array}{c}
\includegraphics[scale=.5]{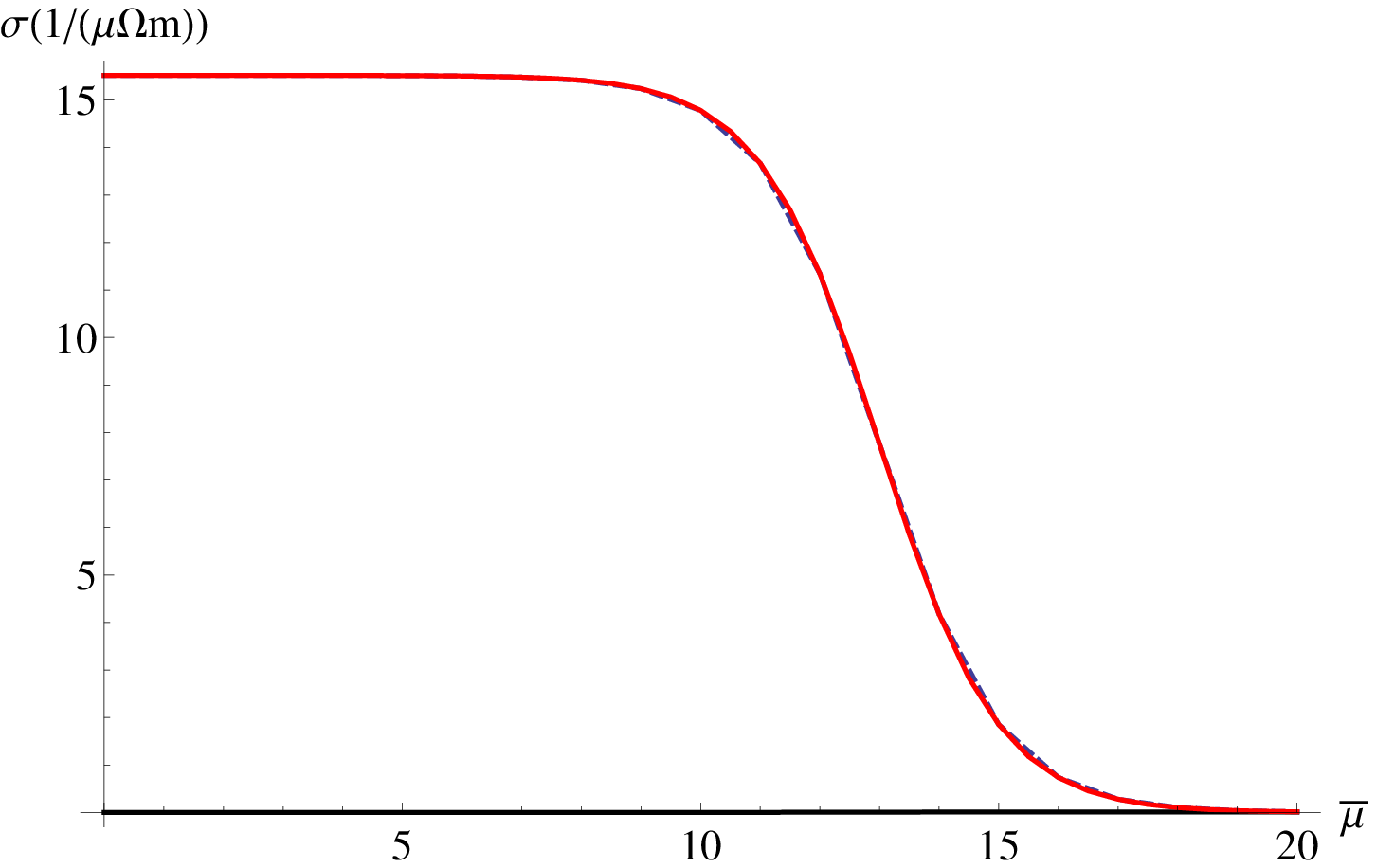}\\
\includegraphics[scale=.5]{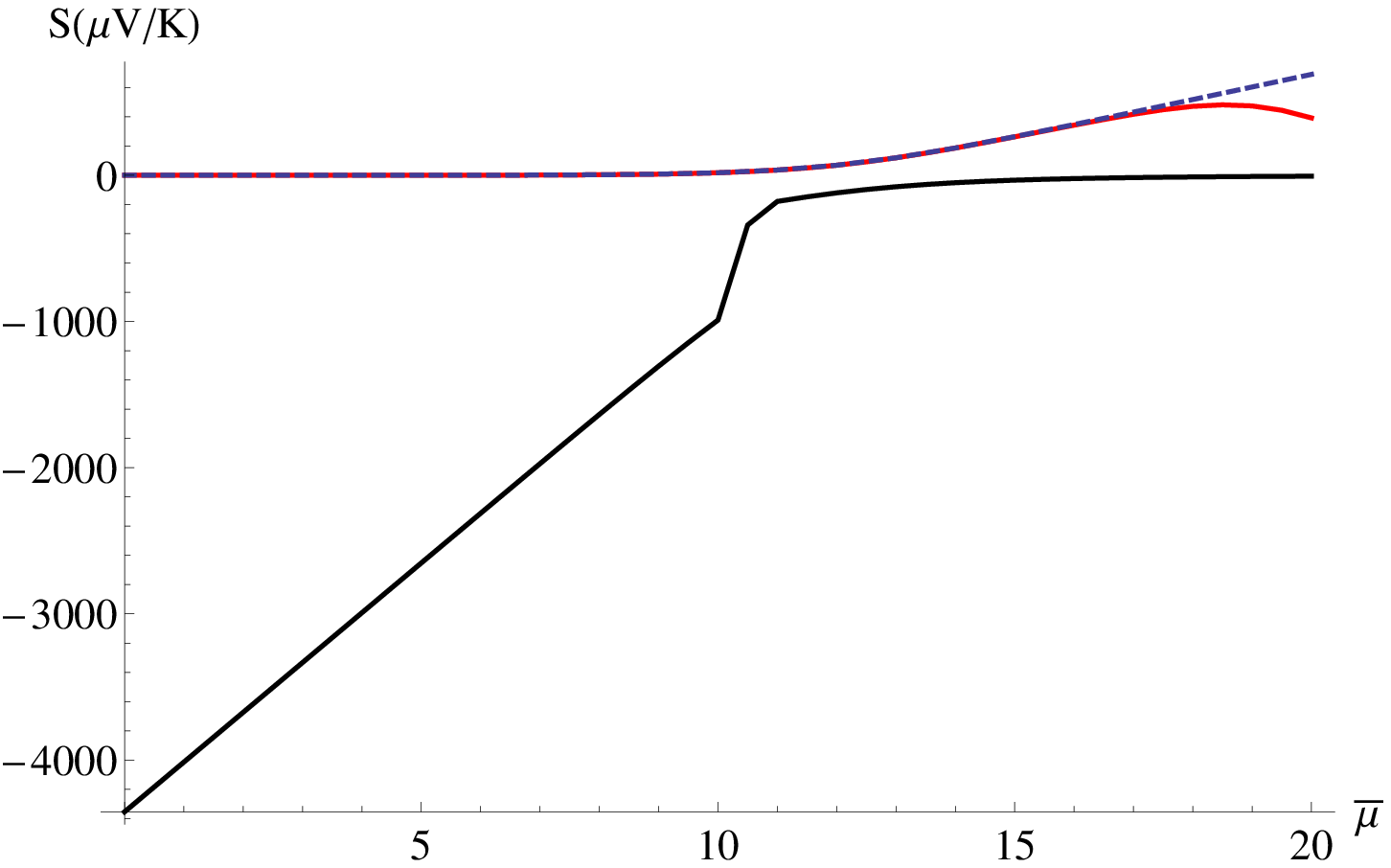} \\
\includegraphics[scale=.5]{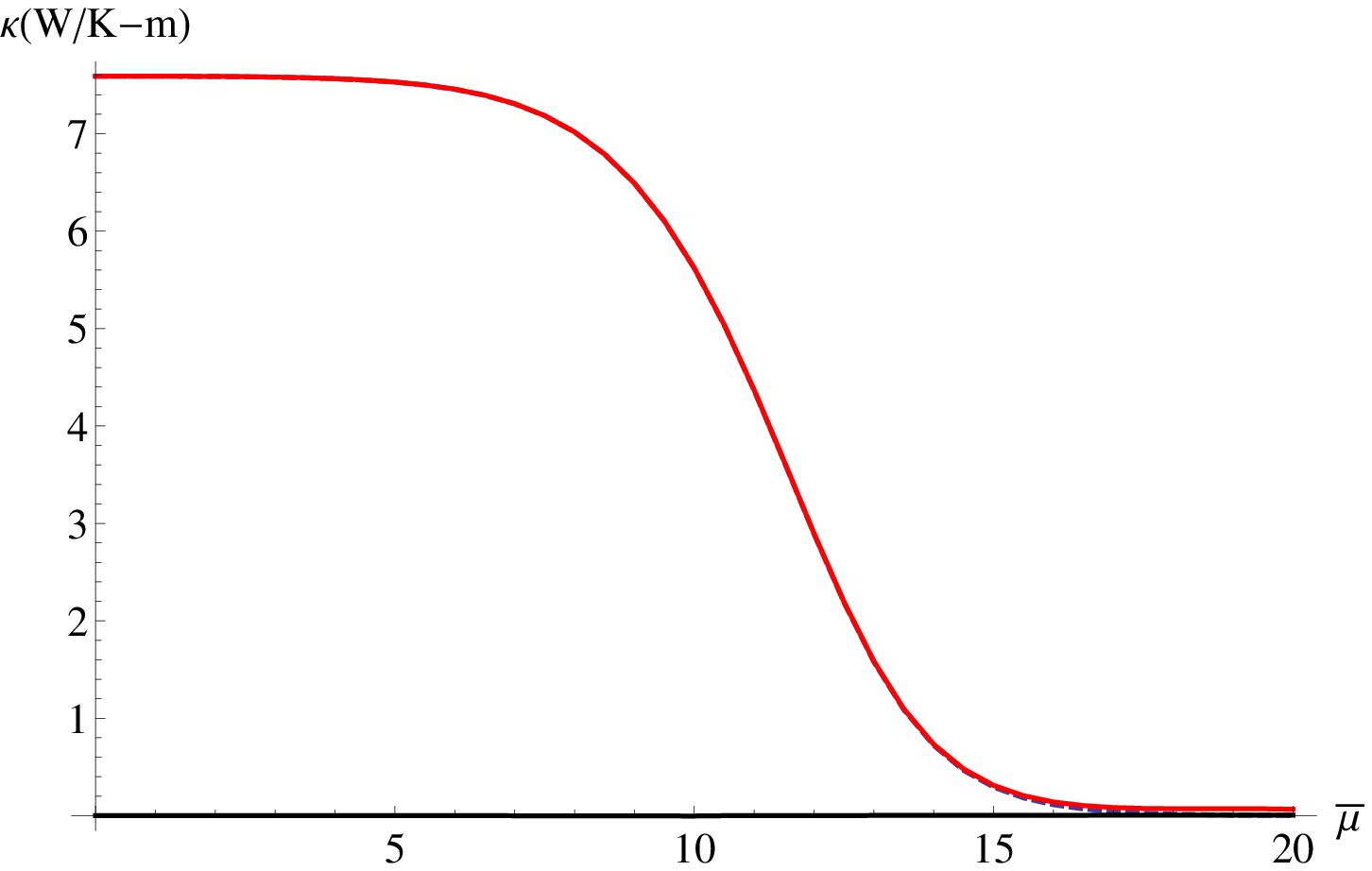}\\
\includegraphics[scale=.5]{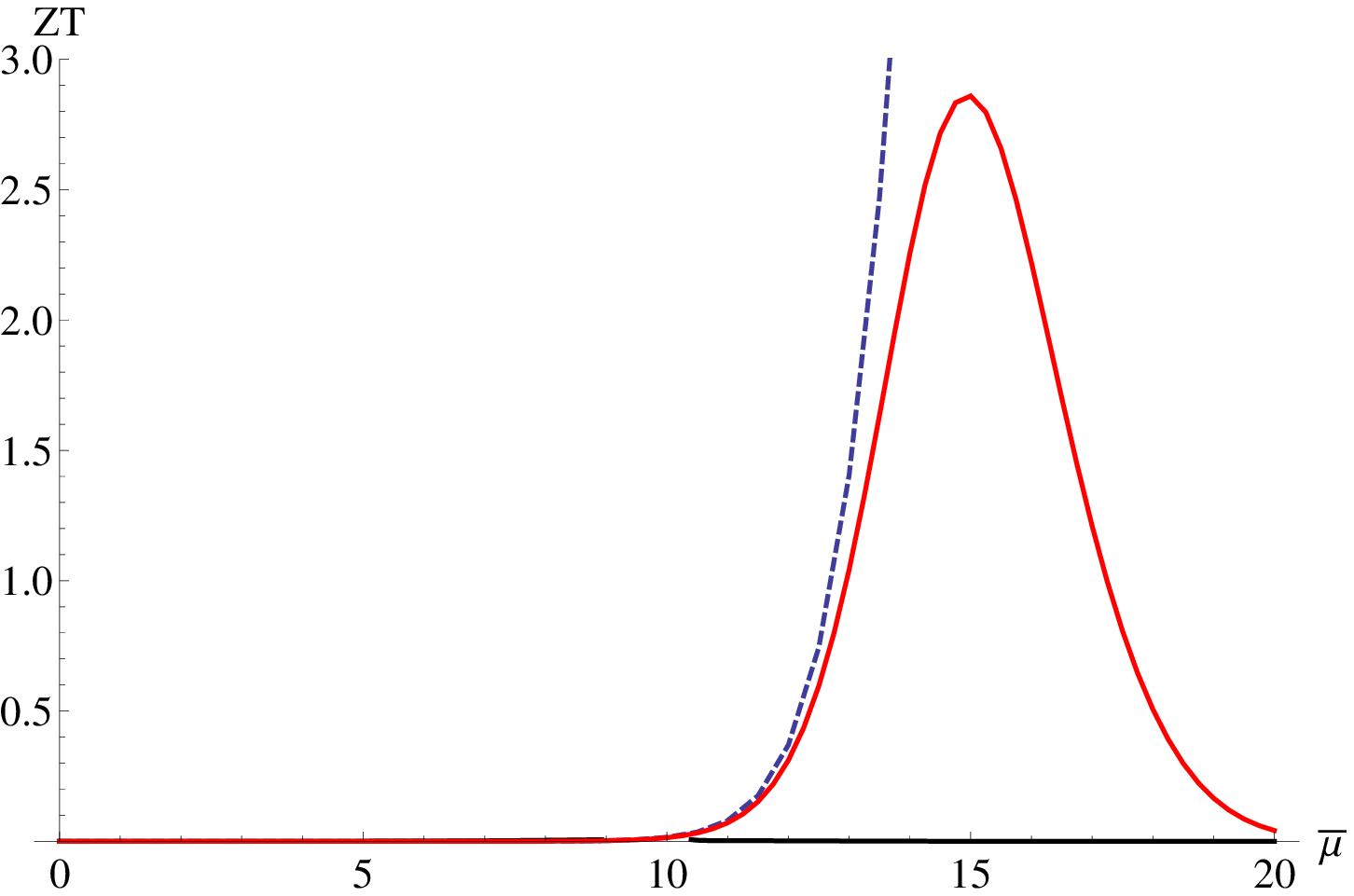}
\end{array}$
\end{center}
\caption{Computation of longitudinal electrical conductivity, Thermopower, electronic 
thermal conductivity and figure of merit as a function of the chemical potential($\mu$)  
at $ T=20K $ for $\ell=1\mu m$. Black curve shows the bulk, dashed curve shows the edge 
and red curve shows the total contribution to the above quantities. For illustration, here, we have chosen $\bigtriangleup_{0}^{1}=\bigtriangleup_{0}^{2}=0.005 z$, $ M_0 =0.075 z$, 
$ t' =-0.3 z$, $\kappa_L=0.5 W m^{-1} K^{-1}$, $ c=0.1 nm$ ($z=0.30$ eV).}
\label{fig.1}
\end{figure}

\begin{figure}[h]
\begin{center}$
\begin{array}{c}
\includegraphics[scale=.5]{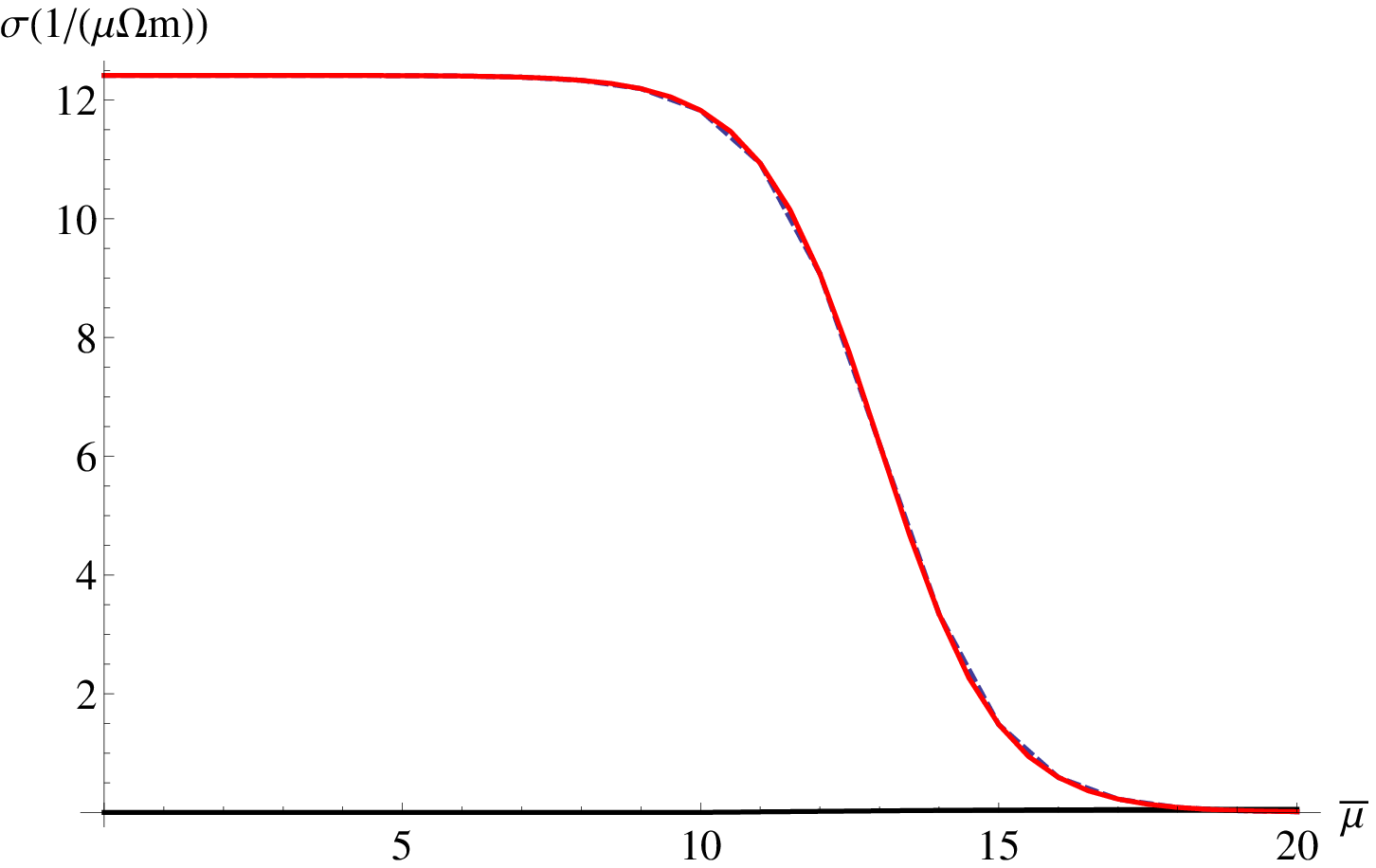} \\
\includegraphics[scale=.5]{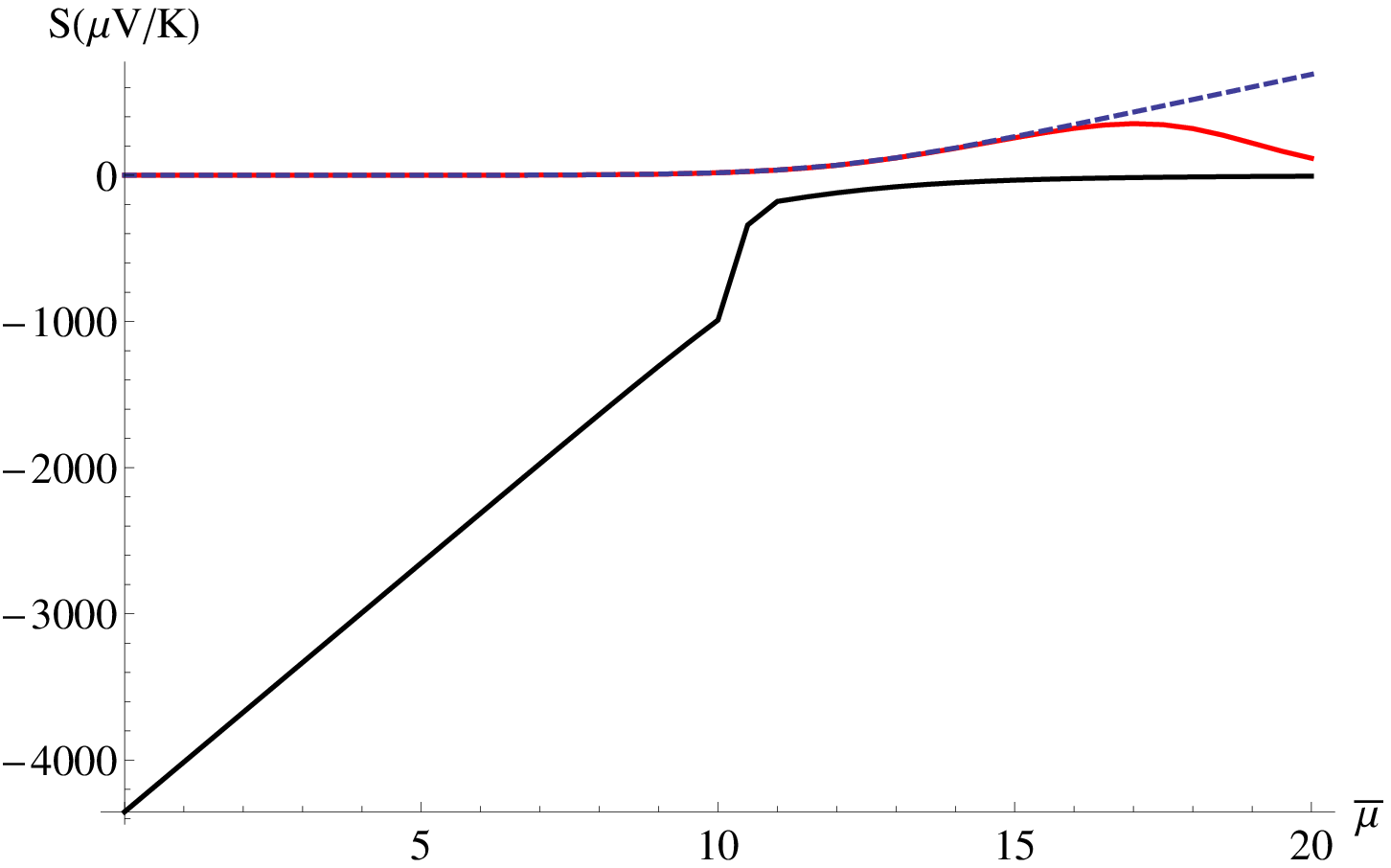}\\
\includegraphics[scale=.5]{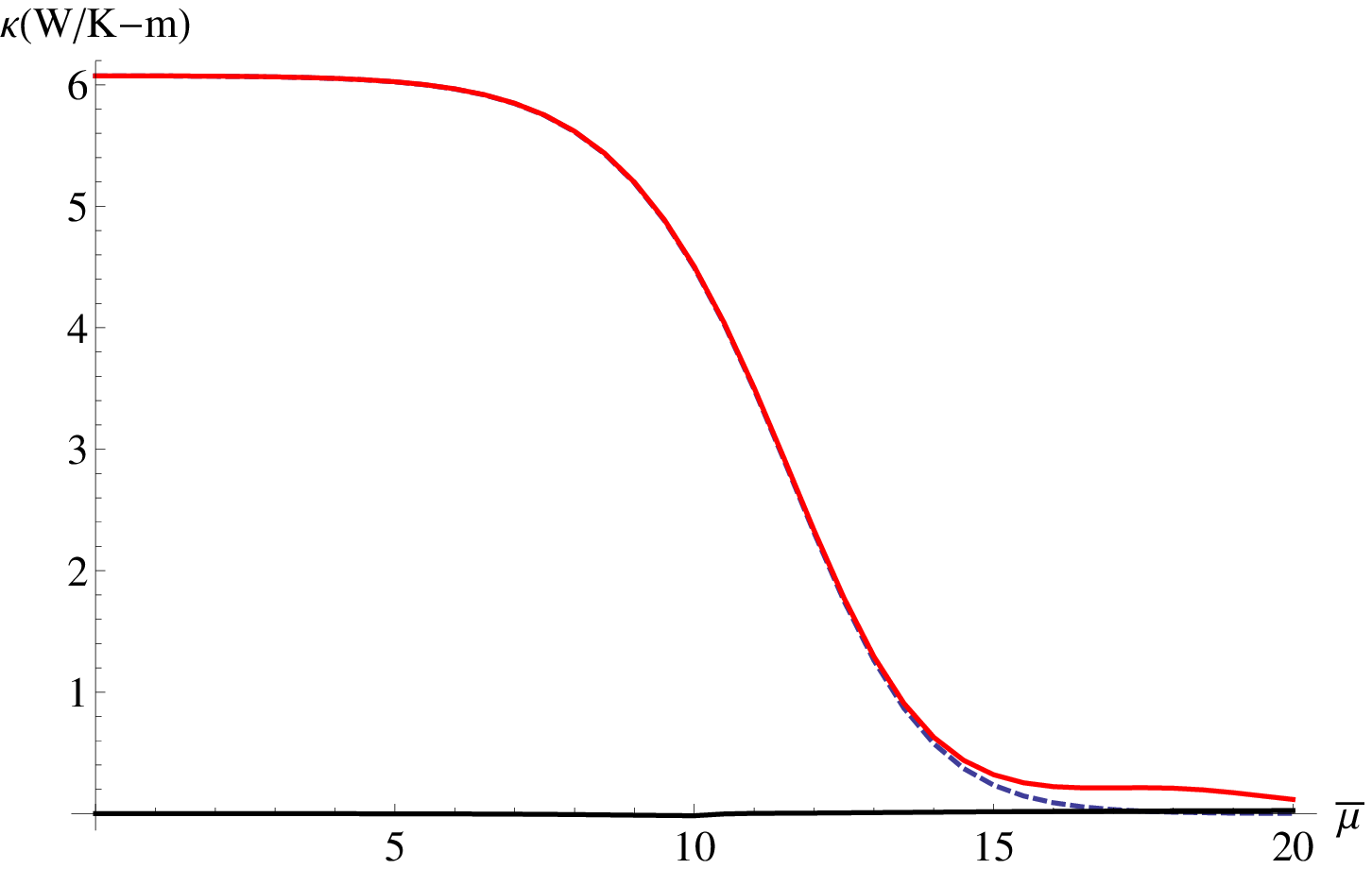}\\
\includegraphics[scale=.5]{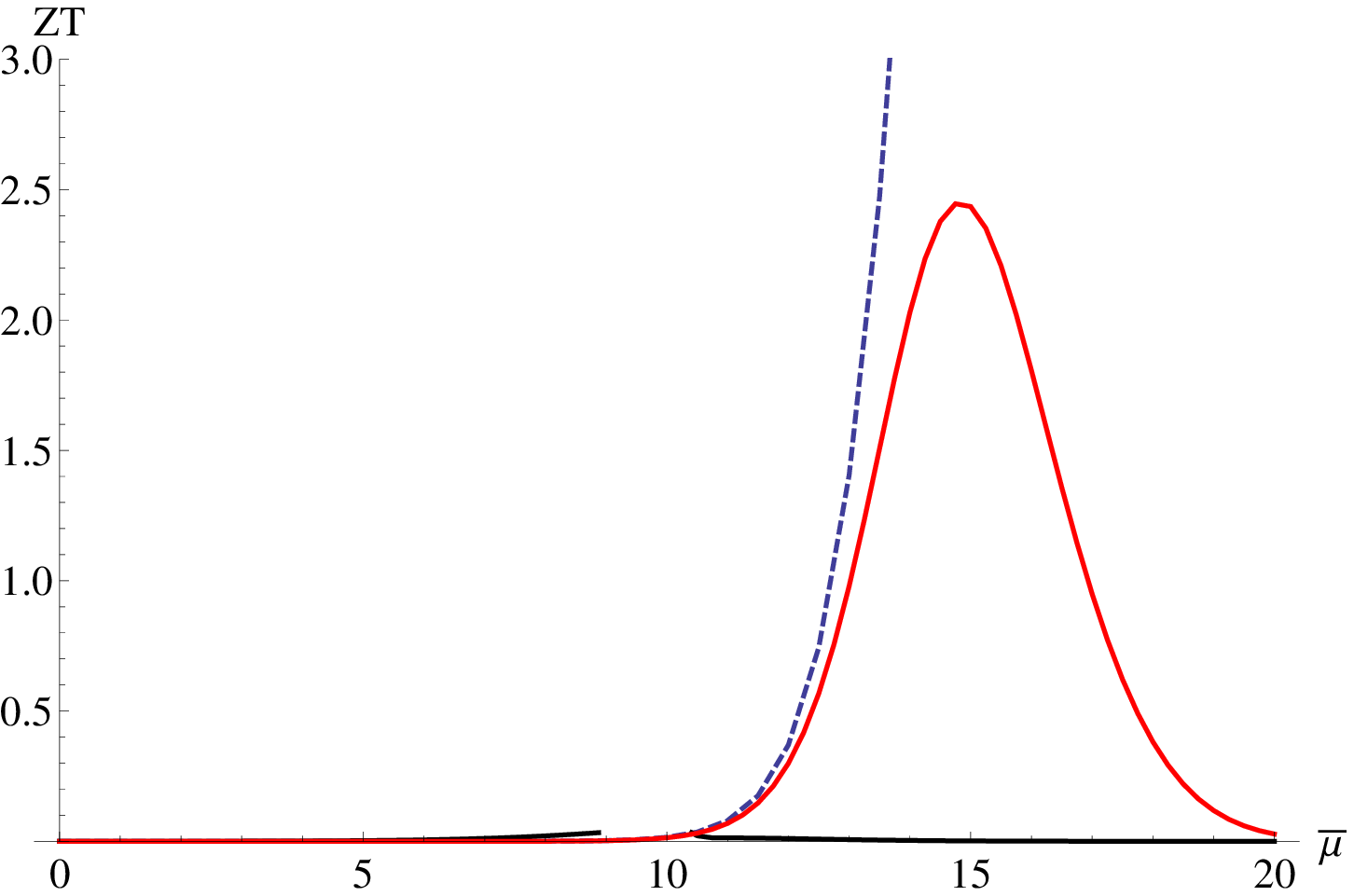}
\end{array}$
\end{center}
\caption{Computation of longitudinal electrical conductivity, thermopower, electronic 
thermal conductivity and figure of merit as a function of the chemical potential($\mu$)  
at $ T=20K $ for $\ell=0.8\mu m$. Black curve shows the bulk, dashed curve shows the edge 
and red curve shows the total contribution to the above quantities. For illustration, here, we have chosen  $ \bigtriangleup_{0}^{1}=\bigtriangleup_{0}^{2}=0.005 z$, $ M_0 =0.075 z$, 
$ t' =-0.3 z$, $ \kappa_L=0.5 W m^{-1} K^{-1} $, $ c=0.1 nm $  ($z=0.30$ eV). }
\label{fig.1}
\end{figure}

 
\end{document}